\input{aipcheck}

\documentclass[
    ,final            
  ]
  {aipproc}

\layoutstyle{8x11double}

\def\gtsima{$\; \buildrel > \over \sim \;$}
\def\ltsima{$\; \buildrel < \over \sim \;$}
\def\simgt{\lower.5ex\hbox{\gtsima}}
\def\simlt{\lower.5ex\hbox{\ltsima}}

\begin{document}

\title{Feedback from the first stars and galaxies and its influence on structure formation}

\classification{<Replace this text with PACS numbers; choose from this list:
                \texttt{http://www.aip..org/pacs/index.html}>}

\author{Benedetta Ciardi}{
  address={Max Planck Institute for Astrophysics; Karl Schwarzschild Str. 1; 85741 Garching; Germany}
}

\begin{abstract}
Once the first sources have formed, their mass deposition, energy
injection and emitted radiation can deeply affect the subsequent
galaxy formation process and influence the evolution of the IGM via a
number of so-called feedback effects. The word 'feedback' is by far one
of the most used in modern cosmology, where it is applied to a vast range
of situations and astrophysical objects. Generally speaking, the concept
of feedback invokes a back reaction of a process on itself or on the
causes that have produced it. The character of feedback can be
either negative or positive. Here,
I will review the present status of investigation of the feedback
effects from the first stars and galaxies.
\end{abstract}

\maketitle


\section{Introduction}

Once the first sources are formed, their mass deposition, energy injection and emitted radiation
can deeply affect the subsequent galaxy and star formation process, and the evolution of the
intergalactic medium (IGM).

Generally speaking, feedback effects can either reduce ({\it negative feedback}) or increase
({\it positive feedback}) the efficiency of the star formation (SF) process.
Although a rigorous classification is not feasible because such effects are produced by the
same sources and it is hard to separate their individual impact,
feedback can be divided into three broad classes (see Ciardi \& Ferrara 2005 for a 
review on the topic): {\it mechanical}, {\it chemical} and {\it radiative}. In addition, reionization
and metal enrichment of the IGM could be analyzed as aspects of feedback, but given their relevance
and the large amount of literature on these subjects, they will be reviewed elsewhere and here I 
will concentrate exclusively on mechanical, chemical and radiative feedback.

\section{Mechanical feedback}

Mechanical feedback is associated with {\it mechanical energy injection from winds and/or SN explosions}.

At low redshift mechanical feedback has been extensively studied both in terms of SN explosions and
galactic outflows. As strong winds from metal-free or extremely metal poor stars are not expected
(although recently few studies demonstrate the contrary; e.g. Meynet, Ekstr\"om, Maeder 2006), 
most of the work at very high 
redshift has concentrated on the effects of SN explosions (Ferrara 1998; Mac Low \& Ferrara 1999;
Nishi \& Susa 1999; Ciardi et al. 2000; Scannapieco, Ferrara \& Broadhurst 2000; Mori, Ferrara \&
Madau 2002; Bromm, Yoshida \& Hernquist 2003; Mackey, Bromm \& Hernquist 2003; Wada \& Venkatesan 2003; 
Salvaterra, Ferrara \& Schneider 2003; Kitayama \& Yoshida 2005; Greif et al. 2007).

\begin{figure}
  \includegraphics[height=.3\textheight]{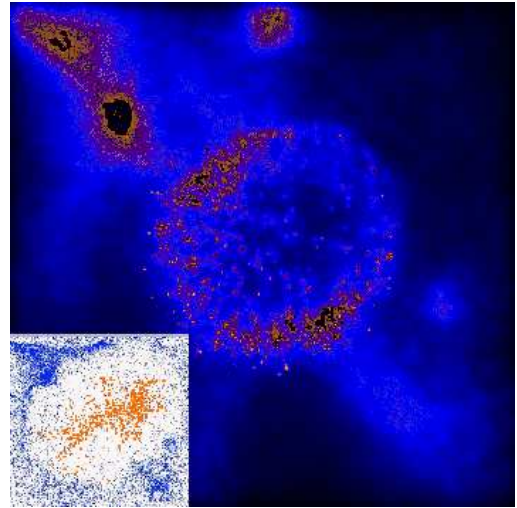}
  \caption{Projected gas density $\sim 10^6$~yr after the explosion of a PISN with mass
$M_\star=250$~M$_\odot$ (corresponding to an explosion energy of $E_{\rm SN}\sim 10^{53}$~ergs)
hosted by a halo of mass $M \sim 10^6$~M$_\odot$ at $z\sim 20$.
The box linear physical size is 1~kpc.
The SN bubble has expanded to a radius of $\sim 200$~pc, having evacuated
most of the gas in the host halo, that has a virial radius of $\sim 150$~pc.
The dense shell has fragmented into numerous cloudlets.
Inset: Metal distribution after 3~Myr. See Bromm, Yoshida \& Hernquist (2003) for details.}
  \label{Bromm03}
\end{figure}

{\it A consequence of SN explosions is to expel the gas out of the host halo and reduce the reservoir
for subsequent star formation}. One of the first studies of such effect from Pair Instability SN (PISN)
is by Bromm, Yoshida \& Hernquist (2003), who run a cosmological SPH simulation until the
conditions for the first star formation are met in a halo of mass $M \sim 10^6$~M$_\odot$ at 
$z\sim 20$. This configuration is then used as initial conditions to follow the effects of the
explosion of such star. The authors assume that the star has either a mass of 
$M_\star=150$~M$_\odot$ (corresponding to an explosion energy of $E_{\rm SN}\sim 10^{51}$~ergs) 
or $M_\star=250$~M$_\odot$ ($E_{\rm SN}\sim 10^{53}$~ergs). They find that in the former case 
the host halo remains almost intact, while in the latter it gets completely disrupted 
(see Fig.~\ref{Bromm03}). 
A similar result has been reached by Greif et al. (2007),
who find that a PISN with $M_\star=200$~M$_\odot$ ($E_{\rm SN}\sim 10^{52}$~ergs) disrupts 
the whole host halo.

\begin{figure}
  \includegraphics[height=.35\textheight]{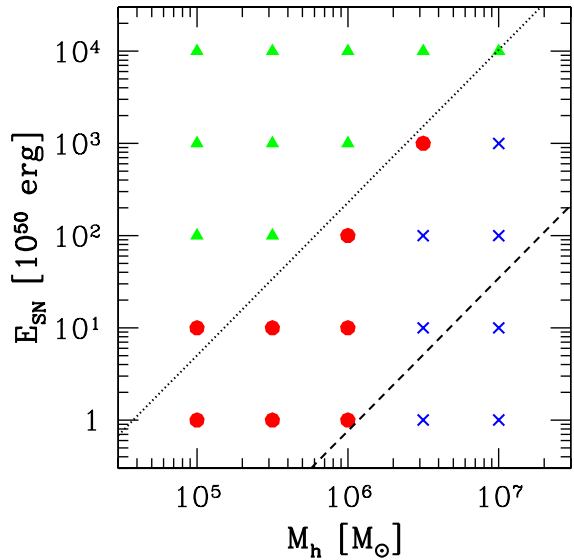}
  \caption{Fate of a halo of mass $M_h$ at $z=20$ hosting a SN explosion with energy $E_{SN}$. Triangles: more than
90\% of the gas is expelled from the halo without pre-existing HII region; circles: more than 90\% of
the gas is expelled from the halo only with pre-existing HII region; crosses: no substantial gas outflow.
Dashed line indicates the halo binding energy (Kitayama \& Yoshida 2005).}
  \label{KitayamaYoshida05}
\end{figure}

A parametric study of the effects of varying the explosion energy and the host halo mass is shown
in Figure~\ref{KitayamaYoshida05}, where the fate of the host halo is characterized by different
symbols. The results have been obtained using a 3D simulation of the formation of HII regions
around the first stars as initial conditions for 1D simulations to follow the effect of SN
explosions (Kitayama \& Yoshida 2005). In general, the authors find that also if the SN energy is larger
than the halo binding energy the gas can still be retained, and that the final fate of the
halo strongly depends on the initial conditions, e.g. the existence and extent of a pre-existing HII region
produced by the SN precursor. Their results are consistent with
the studies mentioned above, as the complete destruction of a halo with mass $M \sim 10^6$~M$_\odot$
follows an explosion with $E_{\rm SN}\sim 10^{53}$~ergs, while the halo remains intact
for $E_{\rm SN}\sim 10^{51}$~ergs.

\begin{figure}
  \includegraphics[height=.55\textheight]{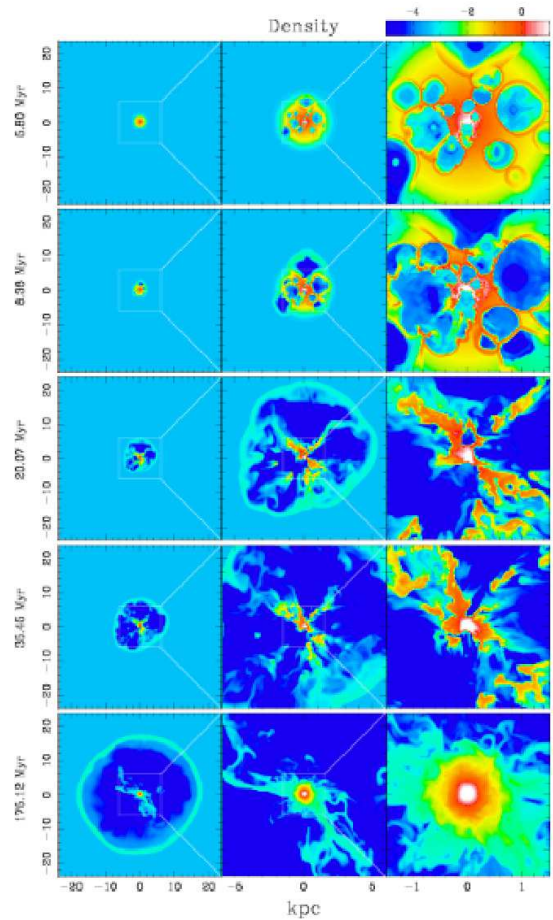}
  \caption{Snapshots of the logarithmic number density of the gas in a halo with
$M=10^8$~M$_\odot$ at $z=9$, at five different elapsed times (from the top 5.80,
8.36, 20.07, 35.45 and 175.12~Myr) after multiple SNae explosions. The three
panels in each row show the spatial density distribution in the $X-Y$
plane on the nested grids (see Mori, Ferrara \& Madau 2002 for details).}
  \label{Mori02}
\end{figure}

The situation is more complicated if instead of a single SN, multiple explosions take place. Although such a case
has been simulated only for a halo with $M=10^8$~M$_\odot$ at lower redshift ($z=9$; Mori, Ferrara \& Madau 2002),
this is a good example of what can be expected at higher $z$. When multiple SNae take place, off-center
explosions drive inward propagating shocks that push the gas to the center where a second episode of SF can
take place (see Fig.~\ref{Mori02}). In this case the situation can be very complex and a positive, rather than negative,
 feedback can be the outcome.

\begin{figure}
  \includegraphics[height=.35\textheight]{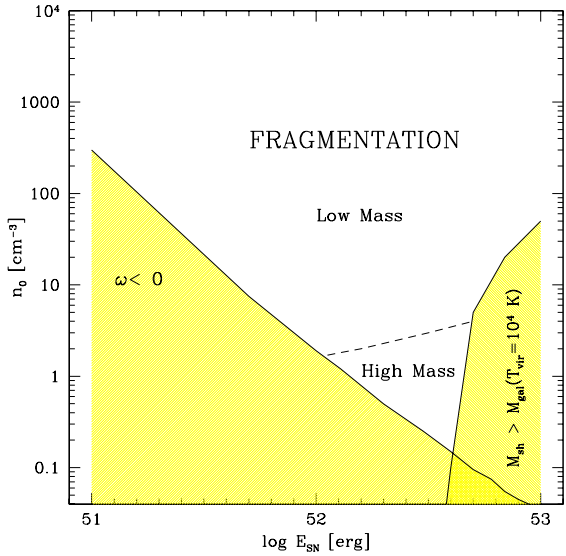}
  \caption{Region of the parameter space $E_{\rm SN}$-$n_0$ where fragmentation into low-mass
($\sim 1$~M$_\odot$) or high-mass ($\sim 100$~M$_\odot$) clumps can occur. Here $n_0$ is the gas density.
Shaded regions are excluded from fragmentation, either because the shell is not gravitationally unstable
($w<0$) or because the shell mass at the time of instability exceeds the total gas mass of the host galaxy
(see Salvaterra, Ferrara \& Schneider 2004 for details).}
  \label{SalvaterraEtal04}
\end{figure}

In addition to this, other positive feedback effects can follow from the propagation of the shocks, which,
sweeping the gas, induce the formation of a dense shell that could eventually fragment and form
stars (Mackey, Bromm \& Hernquist 2003; Salvaterra, Ferrara \& Schneider 2004). Following earlier studies in
local environment, Salvaterra, Ferrara \& Schneider (2004) derived the conditions expected for such 
fragmentation to occur (see Fig.~\ref{SalvaterraEtal04}). Depending on the density of the gas and the
explosion energy, low-mass ($\sim 1$~M$_\odot$) or high-mass ($\sim 100$~M$_\odot$) clumps, and
eventually stars, can form.

\begin{figure}
  \includegraphics[height=.25\textheight]{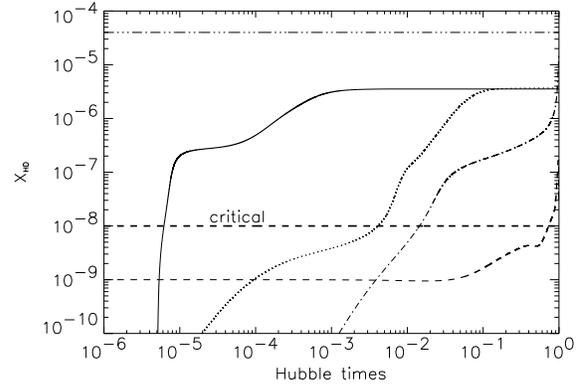}
  \caption{Evolution of the abundance of HD, $X_{\rm HD}$, in primordial gas which cools in four
distinct situations.  The solid line corresponds to gas compressed and heated by a supernova shock at $z=20$.
The dotted line corresponds to gas shocked in the formation of a 3$\sigma$ fluctuation dark matter
halo at $z$ = 15.  The dashed line corresponds to unshocked, un-ionized primordial gas collapsing
inside a mini-halo at $z$ = 20.  Finally, the dash-dotted line shows the fraction of HD in primordial
gas collapsing inside a relic H II region at $z$ = 20. The horizontal line at the top denotes
the cosmic abundance of deuterium.  The critical abundance of HD denoted by the bold dashed line,
is that above which primordial gas can cool to the CMB temperature within a Hubble time. For
details see Johnson \& Bromm (2006).}
  \label{JohnsonBromm06}
\end{figure}

In shocked gas (and in general in a partially ionized medium) HD formation is very efficient. As
HD cooling lowers the temperature of the gas to values below those reached by molecular hydrogen
cooling, the presence of HD in shocks can facilitate fragmentation 
(Vasiliev \& Shchekinov 2005; Johnson \& Bromm 2006; Greif et al. 2007). 
In Figure~\ref{JohnsonBromm06} the HD fraction formed in shocks (solid line) as a function of the Hubble
time is shown, compared to the fraction formed through other mechanisms. It is clear that this is 
by far the most efficient process in producing HD and that its abundance is enough to cool
primordial gas down to the cosmic microwave background (CMB) temperature within a Hubble time (bold dashed line). 

To summarize, the above studies {\it suggest that shocks following SN explosions can induce SF and 
that these stars are typically smaller than those formed in standard conditions}, possibly explaining 
the existence of low-metallicity, low-mass stars. In order {\it to confirm these suggestions 3D studies
are required.}

\section{Chemical Feedback}

Chemical feedback is associated with the {\it existence of a critical metallicity of the gas, $Z_{crit}$, that
induces a transition from a massive to a more standard star formation mode}. After the early work of
Yoshii \& Sabano (1980), chemical feedback has only recently been extensively studied by an
increasing number of authors (Bromm et al. 2001; Schneider et al. 2002; Bromm \& Loeb 2003; Omukai
et al. 2005; Santoro \& Shull 2006; Schneider et al. 2006; Tsuribe \& Omukai 2006; Clark, Glover \&
Klessen 2007; Smith \& Sigurdsson 2007).

\begin{figure}
  \includegraphics[height=.3\textheight]{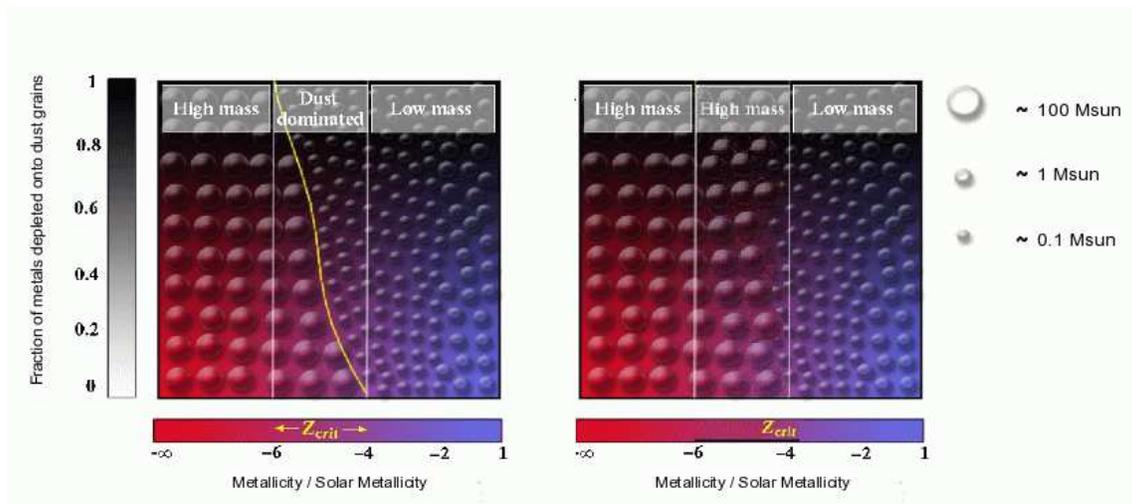}
  \caption{Typical mass of clumps that form from the fragmentation of a gas cloud with different
metallicity and fraction of metals depleted onto dust grains. The case in which dust is (is not)
included is shown in the left (right) panel (original figure in the presence of dust is courtesy
of R. Schneider).}
  \label{chemfeed}
\end{figure}

The fragmentation of gas clouds at various metallicities has been investigated following
different methods: using 1D (Schneider et al. 2002; Bromm \& Loeb 2003; Omukai et al. 2005; 
Santoro \& Shull 2006; Schneider et al. 2006) or 3D (Bromm et al. 2001; Tsuribe \& Omukai 2006; Clark, Glover \&
Klessen 2007) approaches; starting from cosmological initial conditions or focusing on the collapse
of single clouds; including (Clark, Glover \& Klessen 2007) or not rotation of the clouds; etc etc. In the
early works, only H and He chemistry was followed together with metals of varying global metallicity
(Bromm et al. 2001); then the contribution of H$_2$ and of single metal lines (the most important being CII, OI,
SiII and FeII) has been included, and finally also D and dust. What seems to induce the largest scatter in the
results of different investigations is the presence of dust. In fact, studies that include dust 
(Schneider et al. 2002; Omukai et al. 2005; Schneider et al. 2006; Tsuribe \& Omukai 2006;
Clark, Glover \& Klessen 2007) find that clouds with $Z \simlt 10^{-6}$~$Z_\odot$ typically
fragment into clumps of mass $\sim 100$~M$_\odot$, out of which massive stars can collapse, while
if $Z \simgt 10^{-4}$~$Z_\odot$ more typical stars form. In the intermediate range
$Z=10^{-6}-10^{-4}$~$Z_\odot$, the size of
the fragment depends on the amount of metals depleted onto dust grains (left panel of Fig.~\ref{chemfeed}). On the
other hand, in studies where dust is not included (Bromm et al. 2001; Bromm \& Loeb 2003; Santoro \& Shull 2006;
Smith \& Sigurdsson 2007) the formation of high mass stars is prolonged at least to $Z \sim 10^{-4}$~$Z_\odot$
(right panel of Fig.~\ref{chemfeed}).
This is exemplified in Figure~\ref{chemfeed} where the typical clump mass is shown as a function
of the original cloud metallicity and the fraction of metals depleted onto dust grains.
It seems thus {\it crucial to clarify the role and characteristics of dust in the early universe}.

\begin{figure}
  \includegraphics[height=.3\textheight, angle=-90]{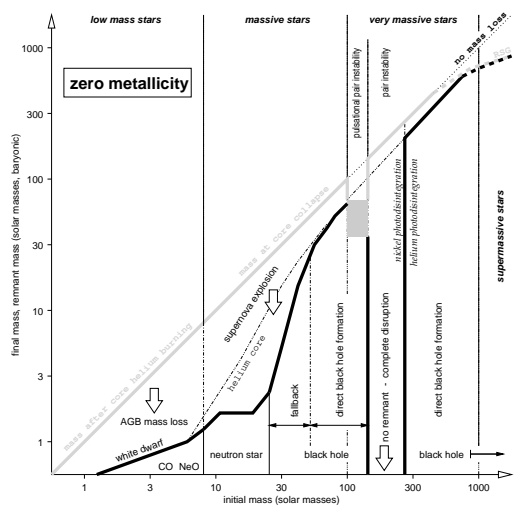}
  \caption{Initial-final mass function of non-rotating metal-free stars.             
The x-axis gives the initial stellar mass. The y-axis
gives both the final mass of the collapsed remnant (thick
black curve) and the mass of the star when the event begins that
produces that remnant (e.g., mass loss in AGB stars, supernova
explosion for those stars that make a neutron star, etc.;
thick gray curve). See Heger \& Woosley (2002) for more details.}
  \label{HegerWoosley02}
\end{figure}

\begin{figure}
  \includegraphics[height=.2\textheight]{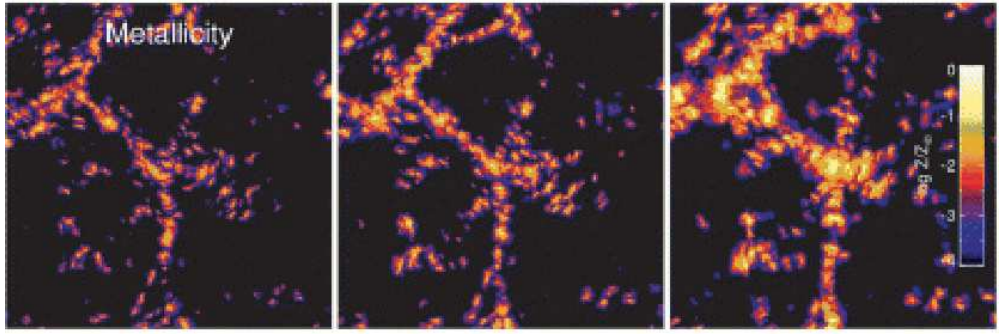}
  \caption{A 100~km~s$^{-1}$ slice showing the gas metallicity at $z=$ 4.5, 3.0, and 1.5, in a
numerical simulation of metal enrichment by Oppenheimer \& Dav\'e (2006).}
  \label{OppenheimerDave06}
\end{figure}

In addition to studying the fragmentation properties of gas clouds with different metallicities,
{\it in order to understand chemical feedback, two additional ingredients are required: the 
initial mass function (IMF) of the first stars and the efficiency of metal enrichment.} 
In fact, only stars in the mass range of SN or PISN explosions contribute to the 
pollution of IGM (see Fig.~\ref{HegerWoosley02}; although, as
mentioned in the previous Section, it seems that also metal-free or extremely metal-poor stars can produce winds and
enrich the surrounding medium), and metal enrichment is
far from being homogeneous also at low redshift, as is both shown in numerical simulations and observations 
(see e.g. Fig.~\ref{OppenheimerDave06}). 

This suggests that the transition from a primordial massive star formation mode towards a more standard
star formation mode, which is highly dependent on the local conditions of the gas, takes place at
different time in different regions of the universe, and thus that the chemical feedback is a local 
process.

\section{Radiative feedback}

Radiative feedback is associated with {\it ionization/dissociation of atoms/molecules and heating
of the gas.}
The literature on this topic is extremely large (Haiman, Rees \& Loeb 1997; Ciardi, Ferrara
\& Abel 2000; Ciardi et al. 2000; Haiman, Abel \& Rees 2000; Susa \& Kitayama 2000; 
Kitayama et al. 2000, 2001; Haiman, Abel \& Madau 2001; Machacek, Bryan \& Abel 2001; 
Ricotti, Gnedin \& Shull 2002; Yoshida et al.
2003; Dijkstra et al. 2004; Shapiro, Iliev \& Raga 2004; Susa \& Umemura 2004; Alvarez, Bromm
\& Shapiro 2006; Ahn \& Shapiro 2007; Ciardi \& Salvaterra 2007; Johnson, Greif \& Bromm 2007)
and an exhaustive summary is a difficult task to achieve.

\begin{figure}
  \includegraphics[height=.55\textheight]{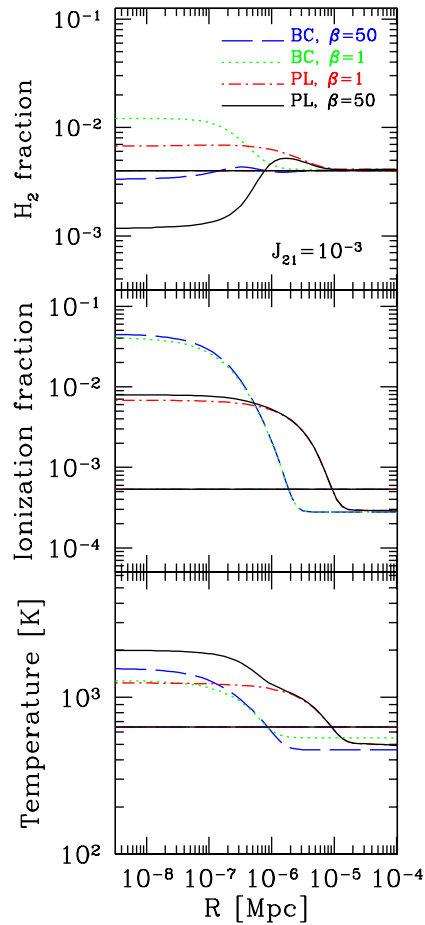}
  \caption{Effect of a point source with a soft (BC) or a hard (PL) spectrum
of different intensity on a uniform medium at different distances from the source. Molecular
hydrogen fraction (top panel), ionization fraction (middle panel) and temperature (lower
panel) are shown (see Ciardi et al. (2000) for details).}
  \label{Ciardi00b}
\end{figure}

The basic concept is that the minimum mass of objects that collapse
in the absence of feedback is $M \sim 10^5$~M$_\odot$.
In the presence of a UV flux instead, objects of such small mass can still form, but
their collapse and the amount of cold gas available for SF depends on the strength of the
feedback. It should be noticed that, while in the early evolutionary stage the radiation
from nearby objects dominates over the background radiation, later on the situation is
inverted (Ciardi et al. 2000).
It has been shown that the intensity and the sign of this feedback depends on a variety of
quantities, like the shape and intensity of the flux. In Figure~\ref{Ciardi00b}
there is a very simple
but illustrative example that shows the effect of a point source with a soft/hard spectrum
of different intensity on a uniform medium at different distances from the source. It can
be seen that, depending on the characteristics of the spectrum and the distance from
the source, either positive 
or negative feedback on molecular hydrogen can be obtained because the main channel for 
the formation of H$_2$ is via H$^-$, and thus the relative abundance of H and electrons
is crucial. The situation gets much
more complicated in more realistic cases, which will be discussed in the following. 

\begin{figure}
  \includegraphics[height=.3\textheight]{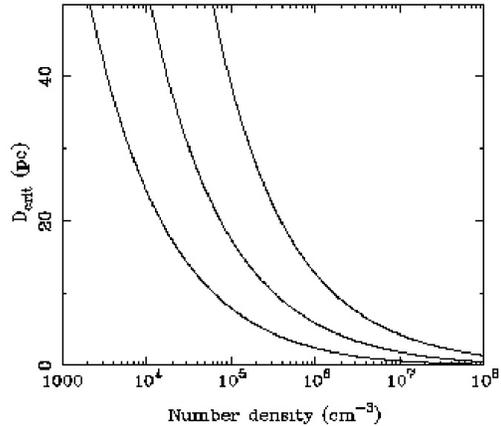}   
  \caption{Distance at which the dissociating time equals the free-fall time as a
function of the cloud number density for different combinations of the H$_2$ abundance and
the mass of the host halo (see Glover \& Brand 2001 for details).}
  \label{GloverBrand01}   
\end{figure}

{\it Once the first generation of stars
has formed in an object, it can affect the subsequent star formation process by dissociating
H$_2$ in nearby star forming clouds}. This is a local, internal feedback. Studies by
Omukai \& Nishi (1999) and Nishi \& Tashiro (2000) show that one massive star produces
enough radiation to dissociate the entire host halo. But, if the
density distribution of the host halo and its geometry are taken into account, {\it SF can proceed 
unimpeded if the star forming clouds are dense and far enough from the star 
that emits the radiation} (Glover \& Brand
2001; Susa \& Umemura 2006). As an illustrative example, in Figure~\ref{GloverBrand01}
the distance at which the dissociating time equals the free-fall time is shown as a
function of the cloud number density for different combinations of the H$_2$ abundance and
the mass of the host halo (see Glover \& Brand 2001 for details). To put the numbers into
perspective, the virial radius of a halo with $M=10^6$ ($10^5$)~M$_\odot$ 
at $z=30$ is about 100 (50)~pc. The result is consistent with a 3D hydro simulation with 
radiative transfer by Susa \& Umemura (2006), who study the effect of a star 
with mass $M_\star=120$~M$_\odot$, finding that clouds further than 30~pc survive
the feedback effect and continue forming stars. This happens not only because of
the absorption and geometrical dilution of the radiation, but also because the shell of H$_2$ forming in
front of the I-front traps the dissociating radiation and shields the cloud. 

\begin{figure}
  \includegraphics[height=.3\textheight]{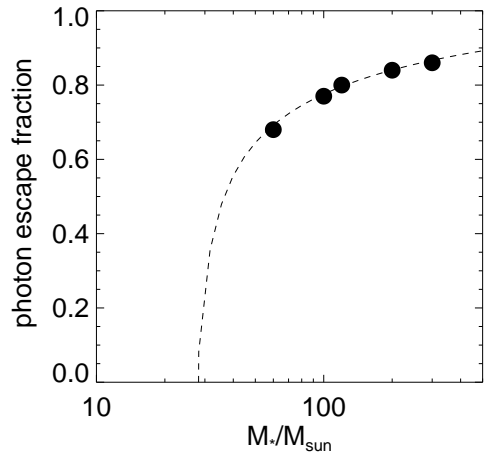}
  \caption{Photon escape fraction as a function of the star mass (Yoshida et al. 2007).}
  \label{Yoshida07}
\end{figure}

In addition to internal feedback, the radiation emitted by the first stars can affects the
IGM and nearby objects evolution. For this reason, it is important to determine
how much of the radiation emitted can escape in the IGM, i.e. the escape fraction
$f_{esc}$. 
Although the situation at low redshift is quite complex and still
controversial (i.e. Dove \& Shull 1994; Ciardi, Bianchi \& Ferrara 2002; Clarke \& Oey 2002;
Fujita et al. 2003), it seems that there is a consensus on the amount of radiation
escaping from massive
stars in the early universe, with {\it $f_{esc}$ being larger than 70\% and increasing with the mass of
the star} (Whalen, Abel \& Norman 2004; Alvarez, Bromm \& Shapiro 2006; Abel, Wise \& Bryan 2007;
Yoshida et al. 2007; but see also Ricotti \& Shull 2000; Wood \& Loeb 2000), as can be seen in 
Figure~\ref{Yoshida07}.

\begin{figure}
  \includegraphics[height=.35\textheight]{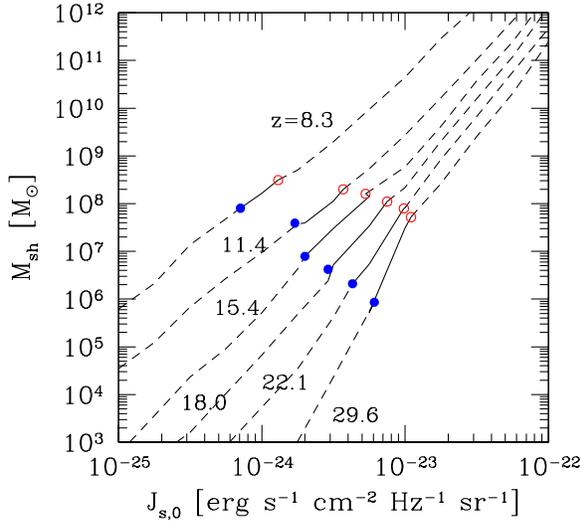}
  \caption{Minimum mass for self-shielding from an external
radiation with varying intensity at  the Lyman limit. Calculations are done for different
redshift. The area affected by the feedback is indicated with solid lines. Objects with
mass larger than that indicated with open circles can cool via H line cooling, while 
filled circles indicate the minimum mass for the collapse in the absence of feedback
(Ciardi et al. 2000).}
  \label{Ciardi00}
\end{figure}

One of the first attempts to model self-consistently several feedback effects is a work
by Ciardi et al. (2000), in which the authors, among others, study, by means of
radiative transfer calculations, the ability
of a halo to self-shield against an external soft-UV radiation and to
collapse. Figure~\ref{Ciardi00} shows the minimum mass for self-shielding from an external
radiation with varying intensity at the Lyman limit. Calculations are done for different
redshifts. The area affected by the feedback is indicated with solid lines.
Subsequent investigations tried to improve the modeling by, e.g., including a proper
treatment of the hydro.

\begin{figure}
  \includegraphics[height=.35\textheight]{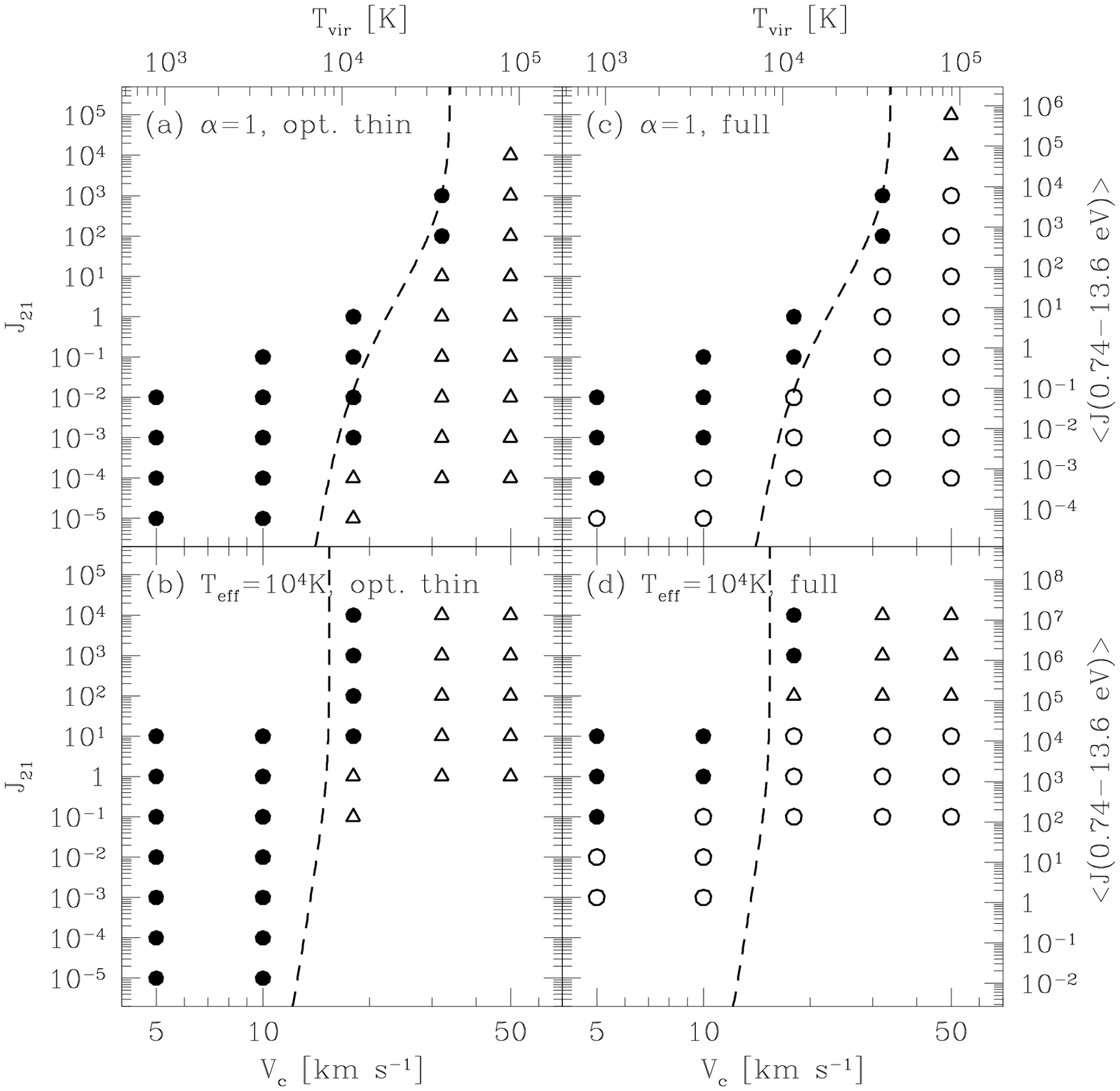}
  \caption{Fate of an object forming in a UV background of varying intensity as a
function of its virial temperature or circular velocity. Full circles: no collapse takes place;
open circles: collapse and cooling take place; open triangles: collapse takes place. 
Different panels refer to different shape of the incident radiation and optically thin
(left panels) or full radiative transfer (right panels) treatment (see Kitayama et al. 2001 for details).}
  \label{Kitayama01}
\end{figure}

The Japanese group of Umemura, Susa, Kitayama and collaborators has performed several studies of the collapse
of a single halo in the presence of a UV background with radiation hydro simulations, both
in 1D and in 3D. In Figure~\ref{Kitayama01} a typical example is shown at $z\sim 10$ for
objects with different virial temperature and incident flux. The results for a stellar
type source with full radiative transfer treatment are summarized in the lower right panel and 
show that if
$J_{21}<0.1$ \footnote{$J_{21}=10^{-21}$~erg~s$^{-1}$~cm$^{-2}$~Hz$^{-1}$~sr$^{-1}$}
all objects collapse and cool, while if the mass of the object is $M \simgt 10^8$~M$_\odot$
collapse and cooling happen independently from the value of $J_{21}$.
Otherwise, their fate depends on $J_{21}$, $M$ and the shape of the spectrum.

\begin{figure}
  \includegraphics[height=.35\textheight]{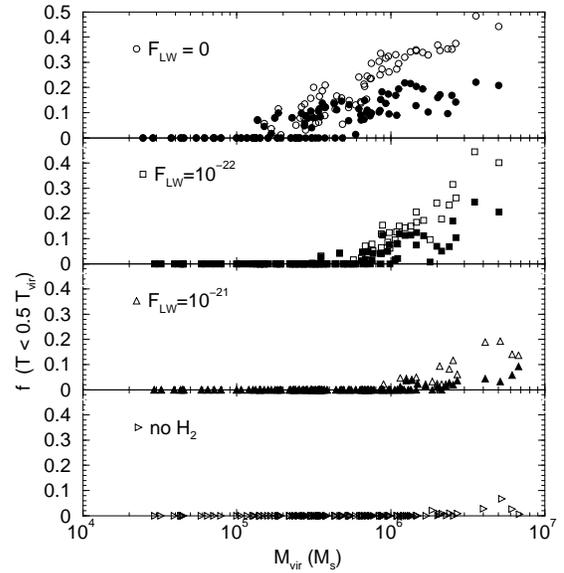}
  \caption{Fraction of cold gas within the virial radius as functions of halo
mass and soft-UV background flux, $F_{LW}$. Open symbols
represent the fraction of gas that has cooled via H$_2$ cooling.
Filled symbols represent the fraction of cold, dense gas available for star formation 
(see Machacek, Bryan \& Abel 2001 for details).}
  \label{Machacek01}
\end{figure}

A different approach is adopted by 
a series of studies that, instead of following the collapse of a single object from
different initial conditions, run cosmological simulations and analyze the fate of the forming halos.
For example, Machacek, Bryan \& Abel (2001) find that the collapse and cooling of small
mass halos is delayed by the presence of a background, and that the amount of cold gas
found in the halos primarily depends on the value of $J_{21}$ and $M$, as can be seen from
Figure~\ref{Machacek01}, in which the fraction of cold gas available for SF in halos in
the redshift range $20<z<30$ is shown as a function of the  
mass of the halos and the background radiation.
This suggests that objects with mass of a
few $10^5$~M$_\odot$ can retain cold gas and form stars if $J_{21}<0.1$.

\begin{figure}
  \includegraphics[height=.6\textheight]{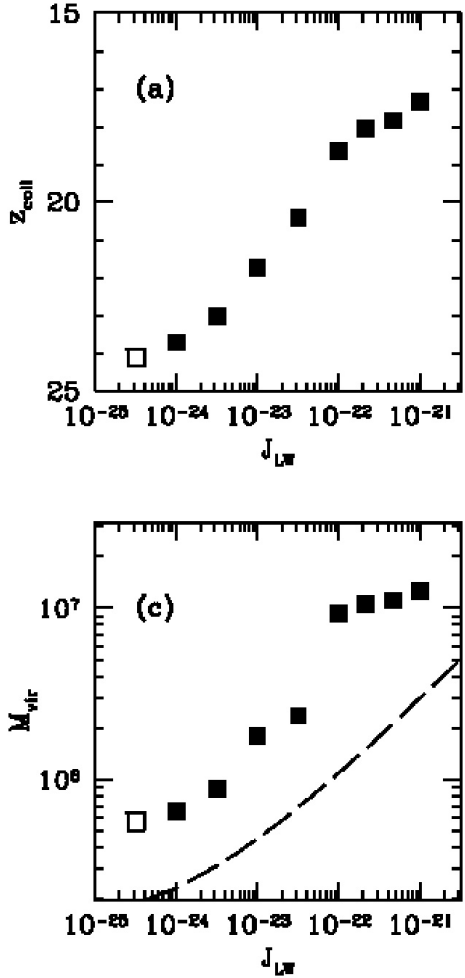}
  \caption{Mean halo quantities for several simulations with the same cosmic
realization but a range of Lyman-Werner molecular hydrogen photodissociating flux
backgrounds (O'Shea \& Norman 2007). Upper panel:  J$_{LW}$ vs. halo collapse redshift.
Lower panel: halo virial mass vs. J$_{LW}$.
The J$_{21} = 0$ ``control'' results are shown as an open square
(it is at log J$_{LW} = -24.5$ in the panels which are a function of J$_{LW}$).
In the bottom panel, the dashed line corresponds to
the fitting function for threshold mass by Machacek, Bryan \& Abel (2001).}
  \label{OSheaNorman07}
\end{figure}

A similar trend (i.e. collapse and cooling are delayed by the presence of a background)
is found also in O'Shea \& Norman (2007). Because of the delay, at the time of the
collapse the objects have actually acquired a larger mass. So, for example, an object that would
collapse at $z=24$ with a mass of $M=5 \times 10^5$~M$_\odot$, in the presence of
a background with $J_{21}=0.1$ would collapse at $z\sim 18$ with a mass $M \sim 10^7$~M$_\odot$
(Fig.~\ref{OSheaNorman07}). This suggests that objects with mass $M<10^6$~M$_\odot$ are not
expected in the presence of a background larger than $J_{21}>0.01$.

Another different approach to the problem is the one by Ahn \& Shapiro (2007). The authors
use a 1D hydro code with radiative transfer to study the effect of the radiation
emitted by a star with mass $M=120$~M$_\odot$ on the formation of halos of different mass 
positioned at different distances from the star.
Although the results are very sensitive to the initial conditions (including the
stage in the halo evolution), objects with $M >10^5$~M$_\odot$ can generally cool and
collapse also if $J_{21}>0.1$ because the I-front gets trapped before reaching the core of 
the halos and the shock that develops in front of it leads to positive feedback. 
The results somewhat contraddict other studies, but they might be affected by the 1D 
configuration which produces 
unphysical geometric effects.
To summarize the effect of dissociating radiation on primordial structure formation
{\it the presence of a UV flux delays the collapse and cooling of small mass halos and the
amount of cold gas available primarily depends on the intensity of the flux
and the mass of the halo. Some constrovery though still exists on the efficiency of such feedback.}

\begin{figure}
  \includegraphics[height=.35\textheight]{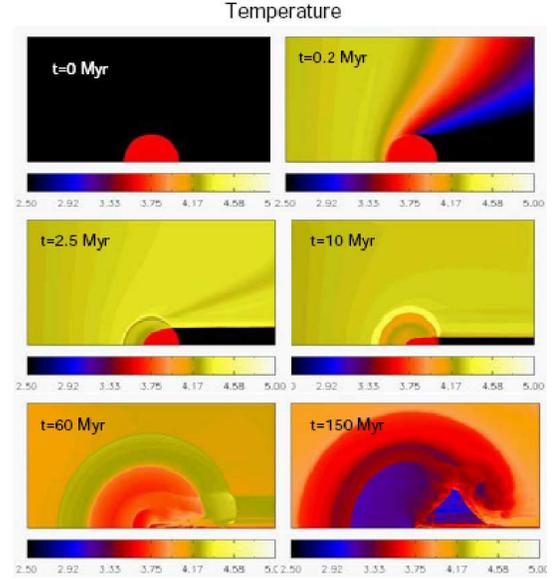}      
  \caption{Temperature evolution of a mini-halo hit by a photoevaporating flux. Different
panel refer to different times (see Shapiro, Iliev \& Raga 2004 for details).}
  \label{Shapiro04}      
\end{figure}

The radiation emitted by the first objects can also 
photoevaporate small mass halos. This effect has been extensively studied e.g. by Shapiro
and collaborators. In Figure~\ref{Shapiro04} a typical example of their simulations
shows the temperature evolution of a mini-halo hit by a photoevaporating flux.
The time scale for complete photoevaporation though is generally longer than the 
lifetime of a PopIII star; thus, {\it in a cosmological context, it is found that mini-halos can
generally survive photoevaporation} (Alvarez, Bromm \& Shapiro 2006; Abel, Wise \& Bryan 2007; but
see also Yoshida et al. 2007). 

\begin{figure}
  \includegraphics[height=.45\textheight]{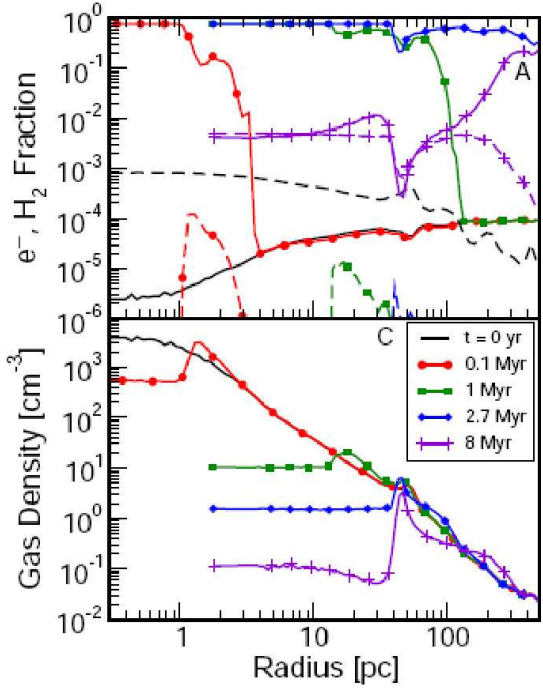}
  \caption{Mass weighted radial profiles around the position of a star with
mass $M_\star=100$~M$_\odot$. Upper panel: electron number 
fraction (solid) and H$_2$ mass fraction (dashed) for 5 different times, 
0, 0.1, 1, 2.7, and 8 Myr after the star is born. The star dies at 2.7~Myr.
Lower panel: density evolution (Abel, Wise \& Bryan 2007).}
  \label{Abel07}
\end{figure}

In addition to the above negative feedback effects, the radiation produced by the
first stars can also favor structure formation, because molecules can efficiently reform, e.g.
inside relic HII regions, and induce a positive feedback 
(Ricotti, Gnedin \& Shull 2001; Nagakura \& Omukai 2005; O'Shea et al. 2005;
Mashchenko, Couchman \& Sills 2006; Abel, Wise \& Bryan 2007; Johnson, Greif \& Bromm 2007;
Yoshida et al. 2007). 
The effect of such positive feedback (that has been first studied by Ricotti
and collaborators) can be seen in Figure~\ref{Abel07}, 
which shows the abundance of molecular hydrogen
and electrons in the vicinity of a source of mass $M_\star=100$~M$_\odot$ at different times. Although
initially H$_2$ is depleted, eventually, once the star has died (in this case after 2.7~Myr), 
it efficiently reforms and is available for a new episode of structure/star formation.

\begin{figure}
  \includegraphics[height=.5\textheight]{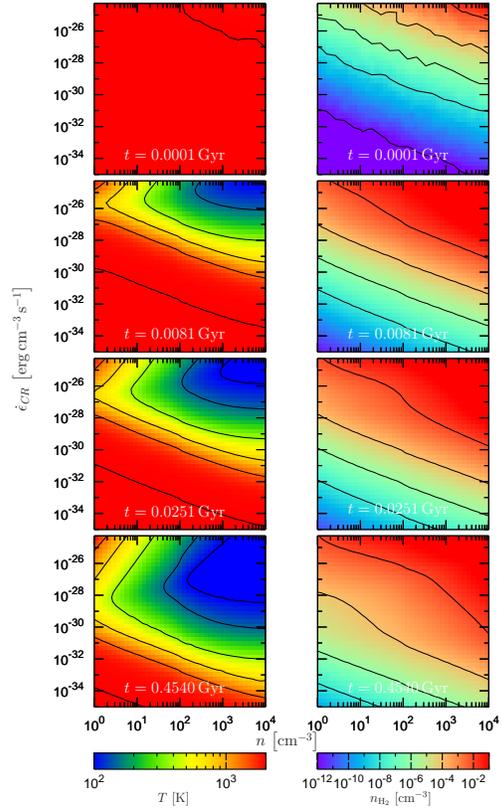}
  \caption{Time evolution of the gas temperature (left panels) and the molecular hydrogen 
abundance (right panels) as a function of the density of the gas, $n$, and of the cosmic 
rays injection rate, $\dot{\epsilon}_{CR}$. The gas is embedded in a dissociating flux 
with $J_{LW}=10^{-21}$~erg~cm$^{-2}$~Hz$^{-1}$~s$^{-1}$.
The superimposed contour plots join points of same temperature or same H$_2$ number density.
See Jasche, Ciardi \& Ensslin (2007) for details.}
  \label{Jasche07}
\end{figure}

In addition to molecular hydrogen, also HD can be efficiently formed inside relic HII regions, with the
consequence of reducing further the gas temperature (Nagakura \& Omukai 2005; Johnson \&
Bromm 2006; Yoshida et al. 2007; Yoshida, Omukai \& Hernquist 2007). 
For this reason, it has been proposed that star formation inside relic HII regions might
produce stars that typically have masses smaller than the stars that have
emitted the radiation.
Note that H$_2$ and HD formation can be further promoted, under certain
conditions, in the presence of x-rays or cosmic rays (e.g. Shchekinov \& Vasiliev 2004; Kuhlen
\& Madau 2005; Jasche, Ciardi \& Ensslin 2007; Stacy \& Bromm 2007). This can be clearly
seen in Figure~\ref{Jasche07}, where the evolution of temperature and molecular hydrogen
is shown as a function of the density of the gas and of the cosmic rays injection rate.

\begin{figure}
  \includegraphics[height=.5\textheight]{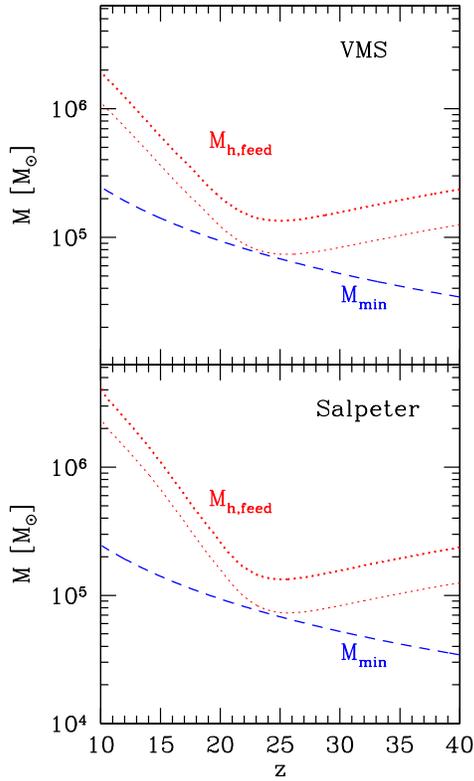}
  \caption{Mass evolution with (dotted lines) and without (dashed lines) Ly$\alpha$ feedback
from a population of metal-free stars with a Salpeter IMF (lower panel) or with mass $M_\star=300$~M$_\odot$.
In objects with mass below $M_{h,feed}$ less than 50\% (and as low as 1\% for the smallest masses) of the
gas is able to collapse compared with a case without feedback.
The two dotted lines refer to different models (see Ciardi \& Salvaterra 2007 for details).}
  \label{CiardiSalvaterra07}
\end{figure}

The reformation mechanism is more efficient than the entropy floor that has been advocated as
a mechanism to prevent accretion and cooling of mini-halos in relic HII regions (Oh \& Haiman
2003; Kramer, Haiman \& Oh 2006). These studies in fact consider adiabatic cooling but, as soon
as some H$_2$ or HD is formed, molecular cooling kicks in and their argument does not
apply anymore.
In order for it to be valid, gas which is heated but not ionized, e.g. by Ly$\alpha$
photons (Ciardi \& Salvaterra 2007), is needed. An example of this can be seen in 
Figure~\ref{CiardiSalvaterra07}, where the effect of Ly$\alpha$ feedback on the ability of the gas to
collapse in the presence of a Ly$\alpha$ background produced by the first, metal-free stars, is shown.

\section{Conclusions}

To conclude, we can summarize feedback effects as follows. In general one can say that:
\begin{itemize}
\item feedback strongly depends on the specific local conditions, e.g. metallicity of the gas,
density of the gas, intensity of ionizing/dissociating radiation etc etc;
\item feedback is not as efficient as naively expected from purely energetic arguments;
\item objects with $M>10^{7-8}$~M$_\odot$ are generally not strongly affected by feedback.
\end{itemize}
Once a primordial star has formed in a halo:
\begin{itemize}
\item star forming regions located in the same halo, but further than a few tens of pc from the star are 
able to shield against its radiation;
\item its UV radiation delays and limit the collapse of cold gas in nearby objects;
\item nearby halos generally survive photoevaporation.
\end{itemize}
After the star dies:
\begin{itemize}
\item formation of H$_2$ and HD in relic HII regions promote structure formation and
low-mass, low-metallicity stars might be the outcome.
\end{itemize}
If the star ends its life as a PISN:
\begin{itemize}
\item the host halo is completely disrupted if its mass is $M<10^6$~M$_\odot$;
\item after the explosion gas in shells can fragment and form small-mass,
low metallicity stars;  
\item metal and dust are expelled and induce a transition to a standard star
formation mode.
\end{itemize}

But several questions still remain at most partially answered, such as: which are
the role and characteristics of dust in the early universe? Which is the IMF of
the first stars? Which is the efficiency of metal enrichment? Is negative feedback
really so negative? Is it possible to have metal-free small-mass stars?




\bibliographystyle{aipproc}   

\bibliography{sample}

\IfFileExists{\jobname.bbl}{}
 {\typeout{}
  \typeout{******************************************}
  \typeout{** Please run "bibtex \jobname" to optain}
  \typeout{** the bibliography and then re-run LaTeX}
  \typeout{** twice to fix the references!}
  \typeout{******************************************}
  \typeout{}
 }

\end{document}